\documentclass[10pt]{article}

\usepackage{PRIMEarxiv}

\usepackage[utf8]{inputenc} % allow utf-8 input
\usepackage[T1]{fontenc}    % use 8-bit T1 fonts
\usepackage{hyperref}       % hyperlinks
\usepackage{url}            % simple URL typesetting
\usepackage{booktabs}       % professional-quality tables
\usepackage{amsfonts}       % blackboard math symbols
\usepackage{nicefrac}       % compact symbols for 1/2, etc.
\usepackage{enumitem}
\usepackage{lipsum}
\usepackage{multirow}
\usepackage[linesnumbered,ruled,vlined]{algorithm2e}
\SetKwInput{Input}{Input}   % Set the Input
\SetKwInput{Output}{Output} % set the Output
\usepackage{amsmath, amssymb, mathrsfs, bbm}
\usepackage{tabularx}
\usepackage{setspace}
\usepackage{subcaption}
\usepackage{xcolor}         % use colors
\usepackage{fancyhdr}       % header
\usepackage{graphicx}       % graphics
\usepackage[backend=biber,style=ieee,sorting=none,natbib=true]{biblatex}
\usepackage{authblk}

\graphicspath{{media/}}     % organize your images and other figures under media/ folder
\usepackage{bm}

\newcommand{\RNum}[1]{\uppercase\expandafter{\romannumeral #1\relax}}

\title{Bilateral Signal Warping for Left Ventricular Hypertrophy Diagnosis}
\author[1,2,*]{Wei Tang}
\author[1,2,*]{Kangning Cui}
\author[2,3]{Raymond H. Chan}
\author[1,$\dagger$]{Jean-Michel Morel}
\affil[1]{ Department of Mathematics, City University of Hong Kong}
\affil[2]{ Hong Kong Centre for Cerebro-Cardiovascular Health Engineering}
\affil[3]{ School of Data Science, Lingnan University}
\date{}                     %% if you don't need date to appear
\begin{document}
\topmargin=0mm

\maketitle

\let\thefootnote\relax
\noindent\footnotetext{$*$ Wei Tang and Kangning Cui contributed equally to this work.}
\noindent\footnotetext{$\dagger$ Corresponding author: jeamorel@cityu.edu.hk}

\begin{abstract}
Left Ventricular Hypertrophy (LVH) is a major cardiovascular risk factor, linked to heart failure, arrhythmia, and sudden cardiac death, often resulting from chronic stress like hypertension. Electrocardiography (ECG), while varying in sensitivity, is widely accessible and cost-effective for detecting LVH-related morphological changes. This work introduces a bilateral signal warping (BSW) approach to improve ECG-based LVH diagnosis. Our method creates a library of heartbeat prototypes from patients with consistent ECG patterns. After preprocessing to eliminate baseline wander and detect R peaks, we apply BSW to cluster heartbeats, generating prototypes for both normal and LVH classes. We compare each new record to these references to support diagnosis. Experimental results show promising potential for practical application in clinical settings.
\end{abstract}

\section{Introduction}
\label{sec:intro}

Left Ventricular Hypertrophy (LVH) is a critical marker of cardiovascular health, characterized by an increase in the mass of the left ventricle in response to chronic stressors. This heart condition is prevalent in people with risk factors such as obesity and diabetes, and also commonly occurs in many patients with untreated hypertension~\cite{kannel1970electrocardiographic, jothiramalingam2021machine, tang2024optimized}. As LVH serves as a strong predictor of cardiovascular diseases, including heart failure, arrhythmia, and sudden cardiac death, early screening and management are essential to control morbidity and mortality~\cite{brady2020electrocardiogram, liu2023left}. 

Electrocardiography (ECG) is a widely accessible and cost-effective screening tool for LVH, despite its varying sensitivity and specificity~\cite{liu2023left, bacharova2014role}. Recently, advances in artificial intelligence (AI) and healthcare data digitization have led to automated LVH diagnostic algorithms using ECG with accuracy on par with expert physicians~\cite{liu2023left, AI_lvh, AI_lvh2}. However, these methods are often criticized as ``black-box'' systems, often lacking interpretability and reliability in clinical settings~\cite{AI1, AI2}. The difficulty in understanding which features drive neural network decisions complicates validation, limits transparency, and poses potential risks to patient safety. This motivates the development of computer-aided methods that can ``mimic'' physician diagnostics and therefore provide supporting evidence for each decision~\cite{interp_ai, interp_ai2}, which has been successfully applied to electroencephalograms\cite{eeg}.

Inspired by Sir William Osler's words, ``\textit{The good physician treats the disease; the great physician treats the patient who has the disease}'', we recognize that doctors often compare patient cases to reach diagnostic conclusions. This raises the question: which patients should serve as prototypes for such comparisons? To solve this, we introduce a bilateral signal warping (BSW) approach in this work to enhance explainable ECG-based LVH diagnosis. Our method constructs a library of heartbeat prototypes by first extracting representative signal patterns for each selected patient candidate. BSW is then applied to group and warp these signals to generate distinct prototypes for normal and LVH cases. These normal and abnormal prototypes act as references to which new ECG records can be compared, thus enabling explainable LVH diagnosis based on both quantitative and visual patient comparison.

\section{Related Works}
\label{sec:related}
\subsection{Left Ventricular Hypertrophy Detection}

The ECG waveform contains identifiable points essential for tracking cardiac cycle phases, including ventricular depolarization~\cite{brady2020electrocardiogram}. In LVH cases, the QRS complex often shows increased amplitude due to the greater electrical load required by hypertrophied ventricular walls. Common voltage-based criteria for LVH diagnosis include the Modified Cornell Criteria, which indicates LVH when the R wave in lead aVL exceeds 1.2 mv, and the Sokolow-Lyon Criteria, where the sum of the S wave in lead V1 and the R wave in lead V5 or V6 surpasses 3.5 mv~\cite{brady2020electrocardiogram, surawicz2008chou}. There exist additional non-voltage indicators, such as a left ventricular ``strain'' pattern marked by ST segment depression and T wave inversion in left-sided leads~\cite{bacharova2014role, T_invert}.

Machine learning methods, including traditional and deep learning approaches, have been used to detect LVH through feature extraction techniques like Fourier or wavelet transforms~\cite{jothiramalingam2021machine}, along with features around the R and S waves to encode ECG signals effectively~\cite{liu2023left}. While these methods reliably differentiate LVH from normal ECG signals, they heavily rely on engineered features --- a contrast to clinical diagnosis, where physicians assess ECG signal patterns directly rather than isolated characteristics.

\begin{figure*}[t]
    \centering
    \includegraphics[width=\linewidth]{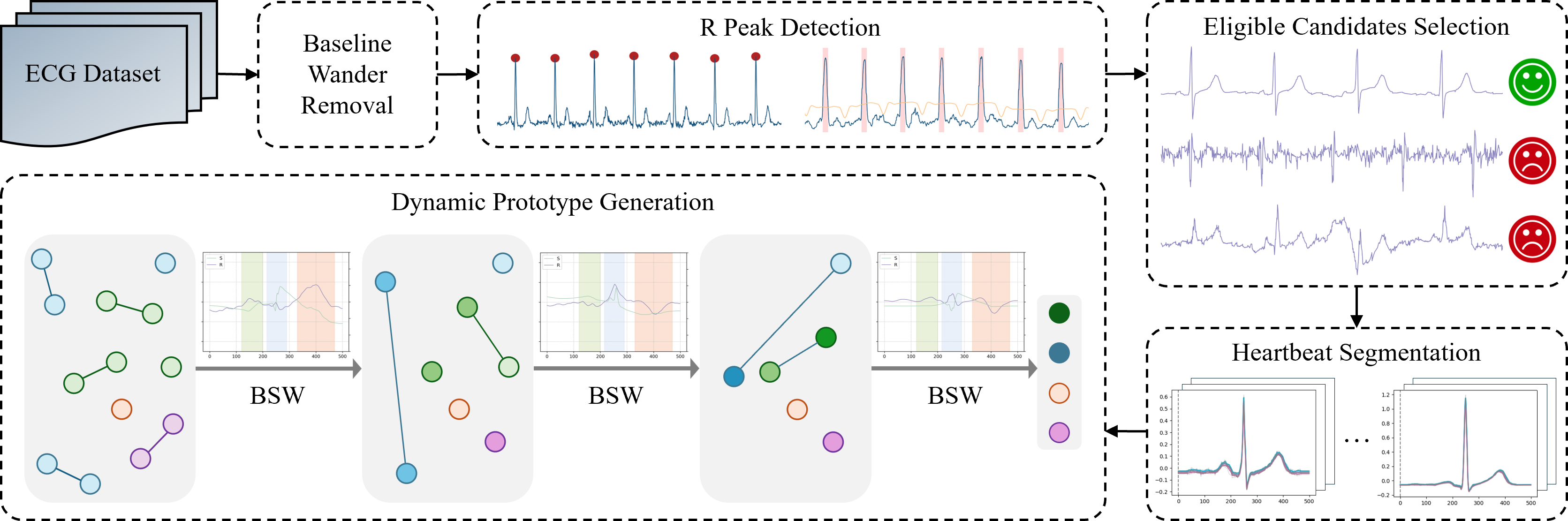}
    \caption{Proposed Prototype Generation Pipeline: The ECG data is preprocessed with baseline wander removal and R peak detection, followed by selection of high-quality signals and computation of mean heartbeats per record. Finally, BSW is applied hierarchically to generate dynamic heartbeat prototypes for the library.}
    \label{fig:pipeline}
\end{figure*}

\subsection{Time Warping}

Time warping methods, especially Dynamic Time Warping (DTW), are widely applied for aligning sequences with temporal variations~\cite{muller2007dynamic}. DTW computes the optimal alignment between two time series $X = \{x_1, x_2, \dots, x_N\}$ and $Y = \{y_1, y_2, \dots, y_M\}$ by minimizing the cumulative distance $D_{i,j}$ over a warping path $W = \{w_1, w_2, \dots, w_K\}$, where each $w_k = (i_k, j_k)$ specifies the alignment between the $i_k$-th element of $X$ and the $j_k$-th element of $Y$. The distance function is recursively defined as:
\[
D_{i,j} = \|x_i - y_j\| + \min(D_{i-1,j}, D_{i,j-1}, D_{i-1,j-1}),
\]
where $D_{i,j}$ represents the cumulative alignment cost at point $(i, j)$ in the two-dimensional grid spanned by $X$ and $Y$. 

In ECG analysis, DTW aligns signals with morphological variability, which enables prototype generation and classification by adjusting for non-linear time shifts. For instance, DTW-based prototypes in~\cite{tziakouri2017classification} classify normal and arrhythmic cases using alignment cost statistics. However, DTW ignores amplitude variations, which is essential for ECG interpretation. Two-dimensional signal warping~\cite{schmidt2014two} addresses this by adjusting time and amplitude jointly, which enhances QT interval variability detection for myocardial infarction analysis. Yet, its non-separable formulation is hardly interpretable. In particular, it restricts explicit differentiation of time and amplitude changes.

In~\cite{morel2023time}, a simple and efficient method is proposed to account for both amplitude and time shifts, using
\begin{equation}
    r(t) f(t) \approx g(t + s(t)),
    \label{eq:warp}
\end{equation} 
where \( r(t) \) adjusts amplitude and \( s(t) \) controls time shifts. Here, \( f(t) \) and \( g(t) \) are the signals to be aligned, and the optimization is achieved through a minimization function, with \( r(t) \) and \( s(t) \) as the target functions. This minimization function $\mathcal{L}(r, s)$ comprises four parts:
\begin{equation}
\begin{aligned}
\mathcal{L}(r, s) = & \int_0^T \left( r(t) f(t) - g(t + s(t)) \right) dt \\ 
& + w_r \int_0^T \left( r'(t) \right)^2 dt + w_s \int_0^T \left( s'(t) \right)^2 dt \\ 
& + w_o \int_0^T \left( s_{\text{min}} - s(t) \right)_+^2 + \left( s(t) - s_{\text{max}} \right)_+^2 dt.
\end{aligned}
\label{eq:loss}
\end{equation}
The first term aligns \( r(t) f(t) \) closely with \( g(t + s(t)) \), accounting for both amplitude scaling and time shifting. The second and third terms encourage smoothness in the functions \( r(t) \) and \( s(t) \), preventing abrupt changes across time. The fourth term constrains \( s(t) \) within \([s_{\text{min}}, s_{\text{max}}]\). Here, \((\cdot)_+\) denotes the ReLU function, which ensures only shifts beyond these bounds contribute to the penalty~\cite{morel2023time}. Initially designed for epidemiological signals, this bilateral signal warping (BSW) method can be reformulated for ECG signals.

\section{Methodology}
\label{sec:methodology}

This section introduces our proposed pipeline for constructing a heartbeat prototype library (see Fig.~\ref{fig:pipeline}). We begin with preprocessing steps to remove baseline wander and detect R peaks for individual heartbeat segmentation. Next, we assess heartbeat variability to select consistent patient data for generating representative prototypes. We then apply BSW to cluster heartbeat prototypes for healthy individuals and LVH patients, with each cluster represented by a signal weighted by its degree of occurrence to capture both common and rare cases. Finally, we validate the library by mapping new patient data to the prototypes, employing BSW statistics to quantify similarity to the closest prototype for diagnostic purposes.

\subsection{Eligible Candidates Selection} 

To construct a reliable heartbeat prototype library, it is essential to only use ``regular'' patients with stable and consistent heartbeats, as irregular or noisy patterns can compromise the integrity of reference prototypes. To systematically identify these candidates, we develop a metric to evaluate ECG signal regularity that allows us to exclude patients with excessive heartbeat variability.

Given an ECG signal with $n$ heartbeats, let $H = \{ h_i(t) \mid i = 1, \ldots, n; t = 1, \ldots, T\}$ represent the set of all heartbeats, where $T$ is the length of each resampled heartbeat. We calculate the mean variability of heartbeats $v(H)$, which captures temporal oscillations across all $h_i(t)$, as follows:
\[
\begin{aligned} 
v(H) &= \frac{1}{T} \sum_{t=1}^T \sqrt{\frac{1}{n-1} \sum_{i=1}^n \left(h_{i}(t) - \hat{h}_t\right)^2}, 
\end{aligned} 
\]
where $\hat{h}_t$ is the mean amplitude at each time step $t$ across all heartbeats. To account for amplitude differences among ECG leads, we normalize $v(H)$ by calculating the average activity $a(H)$:
\[
\begin{aligned}
    a(H) &= \frac{1}{n} \sum_{i=1}^n \sqrt{\frac{1}{T-1} \sum_{t=1}^T \left(h_{i}(t) - \overline{h}_i\right)^2},
\end{aligned}
\]
where $\overline{h}_i$ is the mean value of each heartbeat $h_i(t)$. The final heartbeat variability  
\[
\begin{aligned}
    v_h(H) =: {v(H)}/{a(H)}.
\end{aligned}
\]
serves as a measure of heartbeat consistency across ECG records. By selecting low $v_h$ patients based on a predefined threshold, we can maintain high-quality, uniform heartbeat patterns in the prototype library. The mean heartbeat of these qualified candidates is then utilized in subsequent analyses. This metric supports both initial patient screening and ongoing quality assurance that helps ensure the library’s integrity as new data are integrated.

\subsection{Dynamic Prototype Generation}

We propose bilateral signal warping (BSW) to construct a generalized prototype library from the mean heartbeats of eligible patients. This method improves upon the initial time warping approach, which is restricted to two time series, by addressing non-bipartite matching problems through an iterative process.

Consider a set of n patients $P=\{p_1, ..., p_n\}$, where $p_i$ represents the mean heartbeat of patient $i$, $n = 2^m$, and $m \in \mathbb{N}^*$. We construct an $n\times n$ matrix $D=\{d_{ij}\}$ to quantify the affinity between $p_i$ and $p_j$. This affinity is defined by assessing the variability of $r(t)$ and $s(t)$, derived from the time warping of $p_i$ and $p_j$ by (\ref{eq:warp}). In particular, given that the amplitude of the QRS complex in the ECG signal is a crucial indicator for evaluating LVH, we pay greater attention to the changes of $r(t)$ within the interval containing the QRS complex. The affinity is computed as the summation of the weighted standard deviation of $r(t)$ and the standard deviation of $s(t)$. 

The BSW aims to minimize the total distance between matched elements in each iteration through a non-bipartite matching on $P$. By calculating the reciprocal of each distance, we can utilize the Blossom method~\cite{blossom} to find augmenting paths and the Primal-Dual method~\cite{matchingsurvey} to identify a maximum weight matching, thus determining the most similar heartbeat pairs in each iteration.

\begin{figure}[t]
    \centering
    
    \begin{subfigure}[t]{\linewidth}
    \centering
    \includegraphics[width = \textwidth]{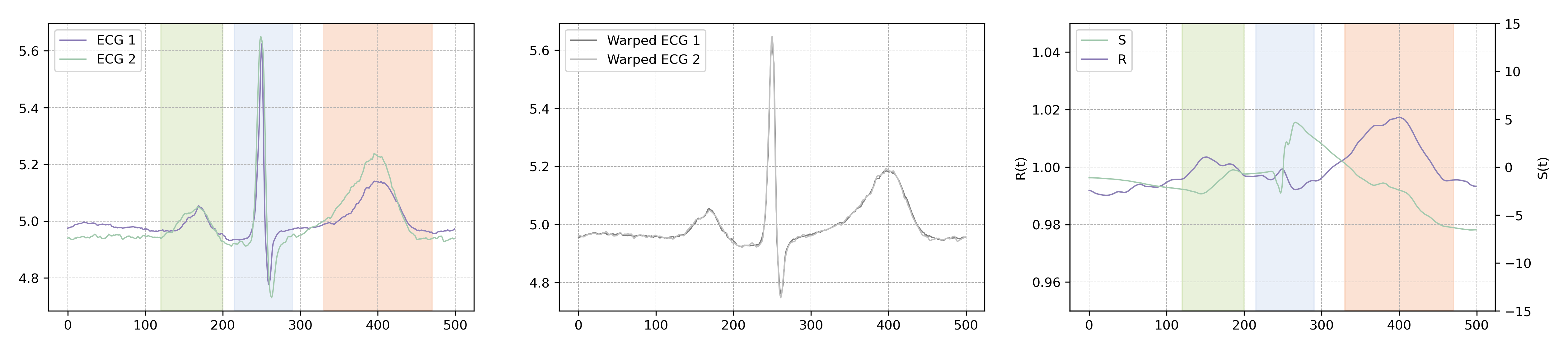} \hspace{0.02in}
    \vspace{-0.5cm}
    \caption{Lead \RNum{2}}
    \end{subfigure}
    
    \begin{subfigure}[t]{\linewidth}
    \centering
    \includegraphics[width = \textwidth]{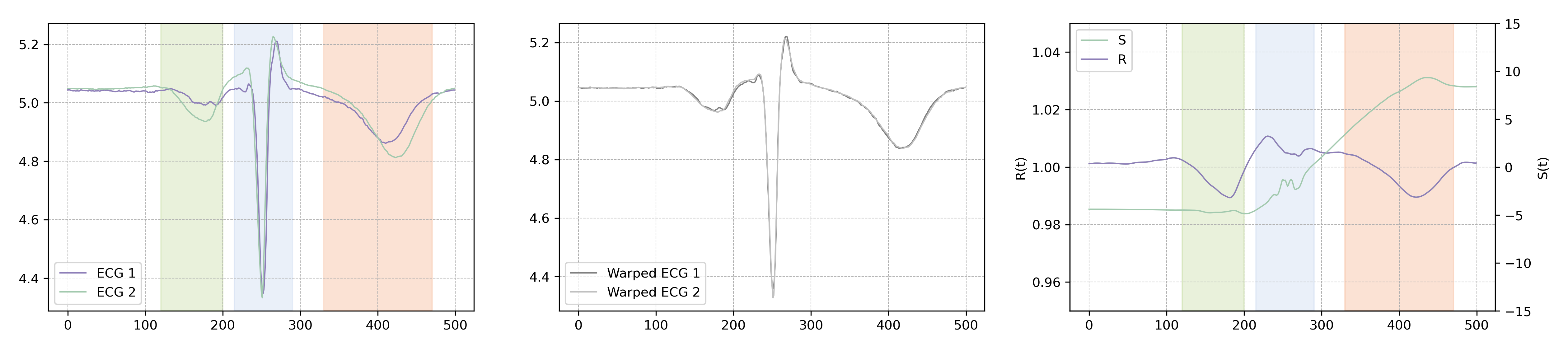} \hspace{0.02in}
    \vspace{-0.5cm}
    \caption{Lead aVR}
    \end{subfigure}
    
    \vspace{-0.2cm}
    \caption{Time Warping of ECG Heartbeats: The first two columns display the ECG signals before and after warping, while the last column presents the output $r(t)$ and $s(t)$. The shaded areas emphasize the P, R, and T waves.}
    \label{fig:warp}
\end{figure}

Recognizing that each ECG lead may contain multiple heartbeat prototypes, we set thresholds for $r(t)$ and $s(t)$. If the thresholds are satisfied, we warp the pair and use $\frac{1}{2}(\sqrt{r(t)} f(t-\frac{s(t)}{2})+ \frac{1}{\sqrt{r(t)}}g(t+\frac{s(t)}{2}))$ as the warped heartbeat for the next iteration. Conversely, if a heartbeat lacks an optimal match, it is designated as a potential new prototype or deemed unsuitable for warping in the current iteration, enabling it to proceed unaltered to the next round. Ultimately, the process converges, yielding the desired prototype library for a typical lead, with the degree of occurrence for each prototype determined by the number of patients contributing to its creation.

\section{Experiments}
\label{sec:experiments}

\begin{figure*}[t]
    \centering
    
    \begin{subfigure}[t]{0.49\textwidth}
    \centering
    \includegraphics[width = \textwidth]{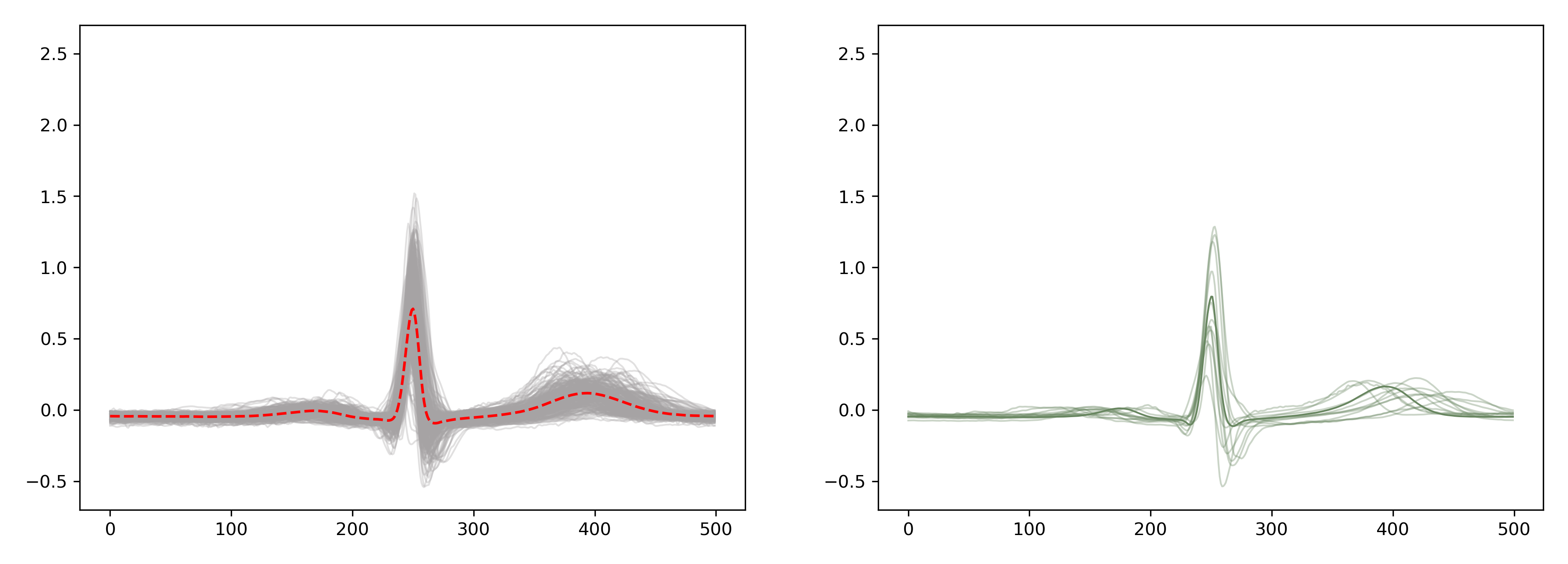} \hspace{0.02in}
    \vspace{-0.5cm}
    \caption{Lead \RNum{1}, Normal}
    \end{subfigure}
    \begin{subfigure}[t]{0.49\textwidth}
    \centering
    \includegraphics[width = \textwidth]{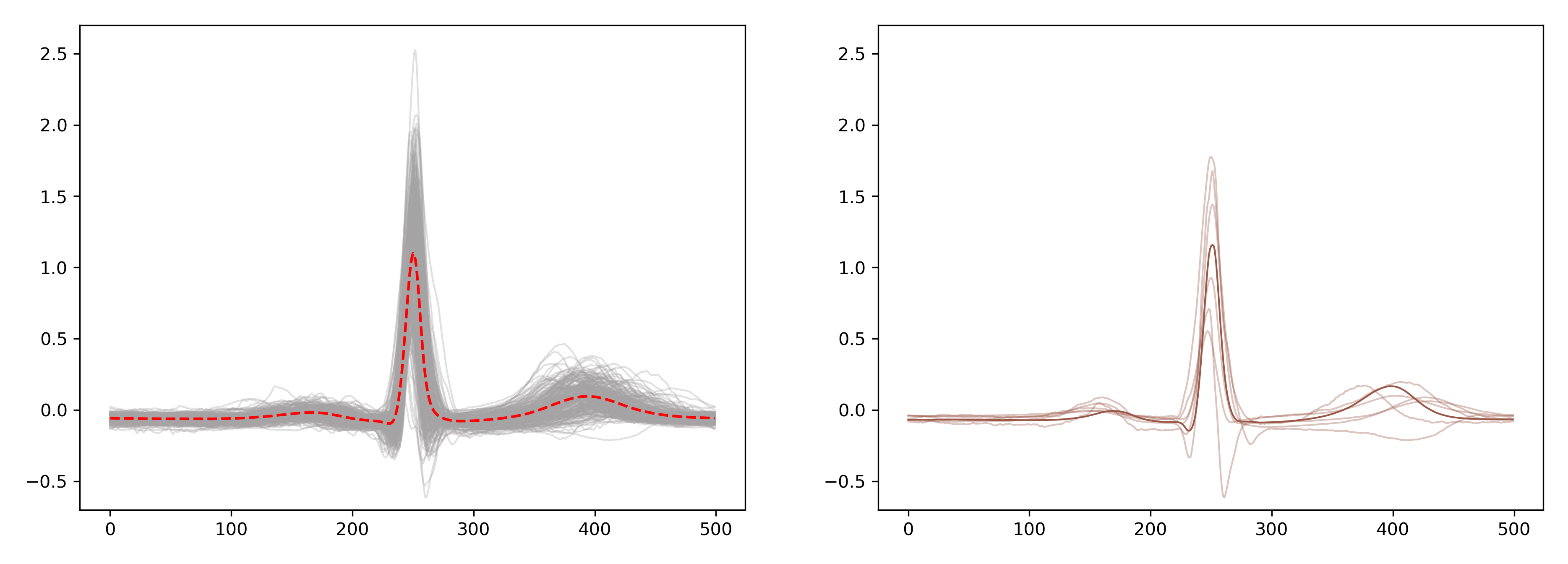} \hspace{0.02in}
    \vspace{-0.5cm}
    \caption{Lead \RNum{1}, LVH}
    \end{subfigure}
    
    \begin{subfigure}[t]{0.49\textwidth}
    \centering
    \includegraphics[width = \textwidth]{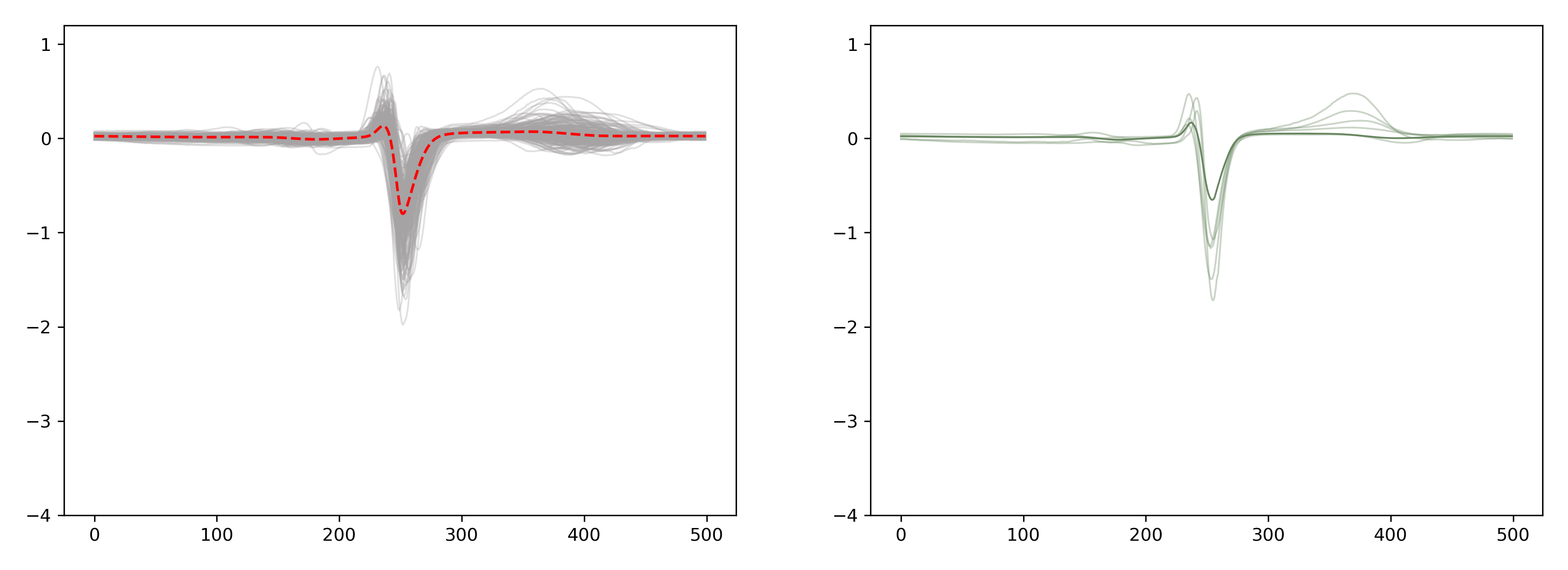} \hspace{0.02in}
    \vspace{-0.5cm}
    \caption{Lead V1, Normal}
    \end{subfigure}
    \begin{subfigure}[t]{0.49\textwidth}
    \centering
    \includegraphics[width = \textwidth]{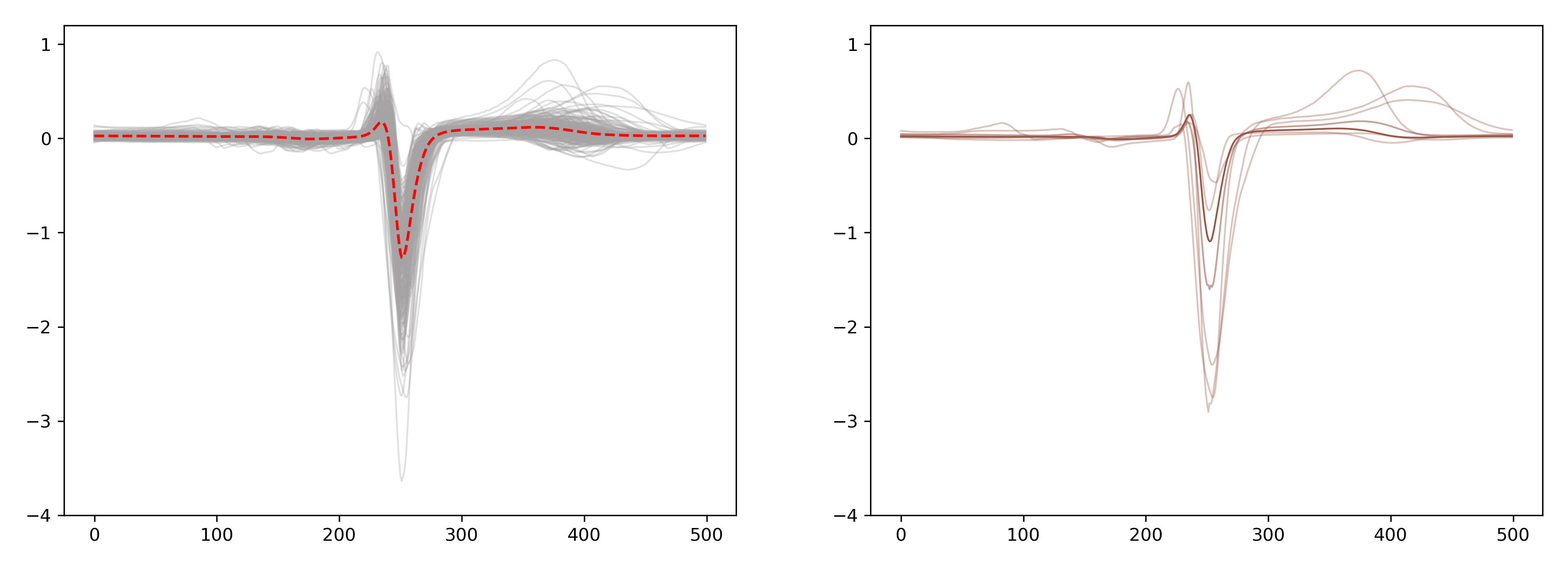} \hspace{0.02in}
    \vspace{-0.5cm}
    \caption{Lead V1, LVH}
    \end{subfigure}
    
    \begin{subfigure}[t]{0.49\textwidth}
    \centering
    \includegraphics[width = \textwidth]{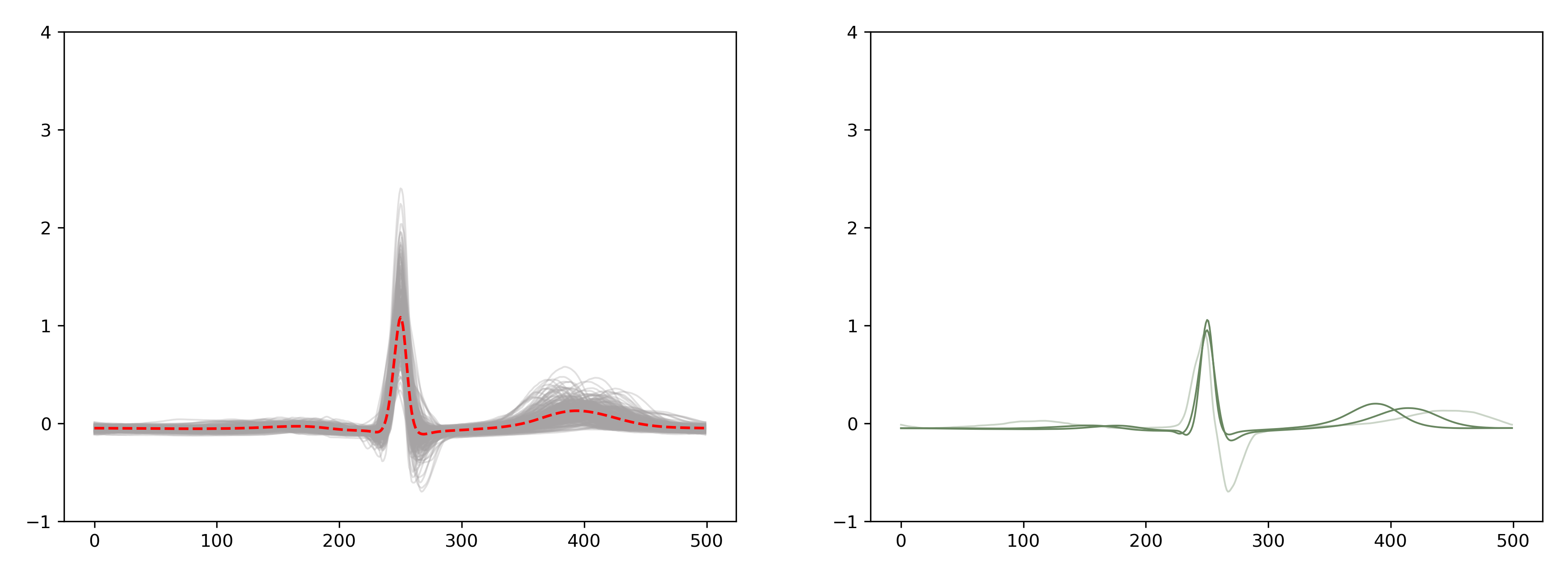} \hspace{0.02in}
    \vspace{-0.5cm}
    \caption{Lead V6, Normal}
    \end{subfigure}
    \begin{subfigure}[t]{0.49\textwidth}
    \centering
    \includegraphics[width = \textwidth]{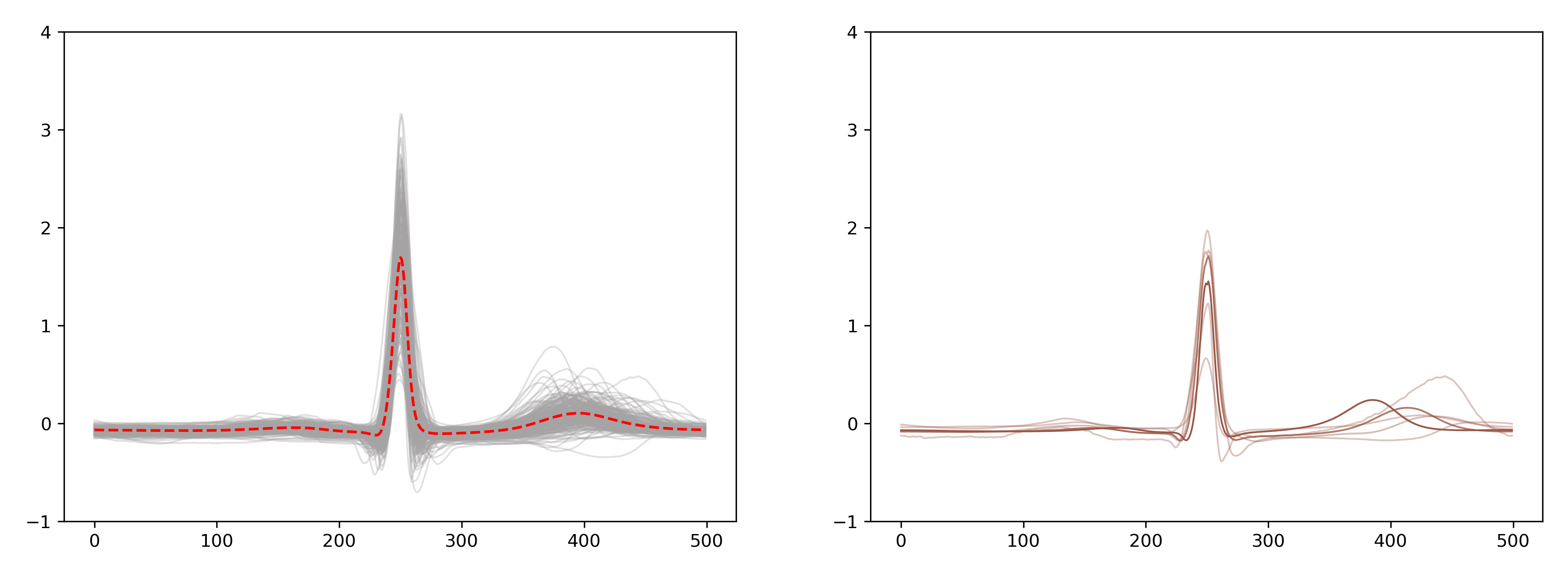} \hspace{0.02in}
    \vspace{-0.5cm}
    \caption{Lead V6, LVH}
    \end{subfigure}
    
    \vspace{-0.2cm}
    \caption{Heartbeat Prototypes for Normal and LVH from Lead \RNum{1}, V1, and V6. Each panel presents two subfigures: Left figures show individual heartbeats (grey) with averaged waveforms (red), while right figures display BSW-generated prototypes, where darker colors indicate higher occurrence rates.}
    \label{fig:comparison}
\end{figure*}

% \begin{figure}[ht]
%     \centering
    
%     \begin{subfigure}[t]{\linewidth}
%     \centering
%     \includegraphics[width = \textwidth]{figures/confusion_m.png} \hspace{0.02in}
%     \vspace{-0.5cm}
%     \caption{Lead V6}
%     \end{subfigure}
    
%     \vspace{-0.2cm}
%     \caption{1: LVH, 0: NORMAL.}
%     \label{fig:confmatrix}
% \end{figure}

\begin{figure}[ht]
    \centering
    \includegraphics[width = \linewidth]{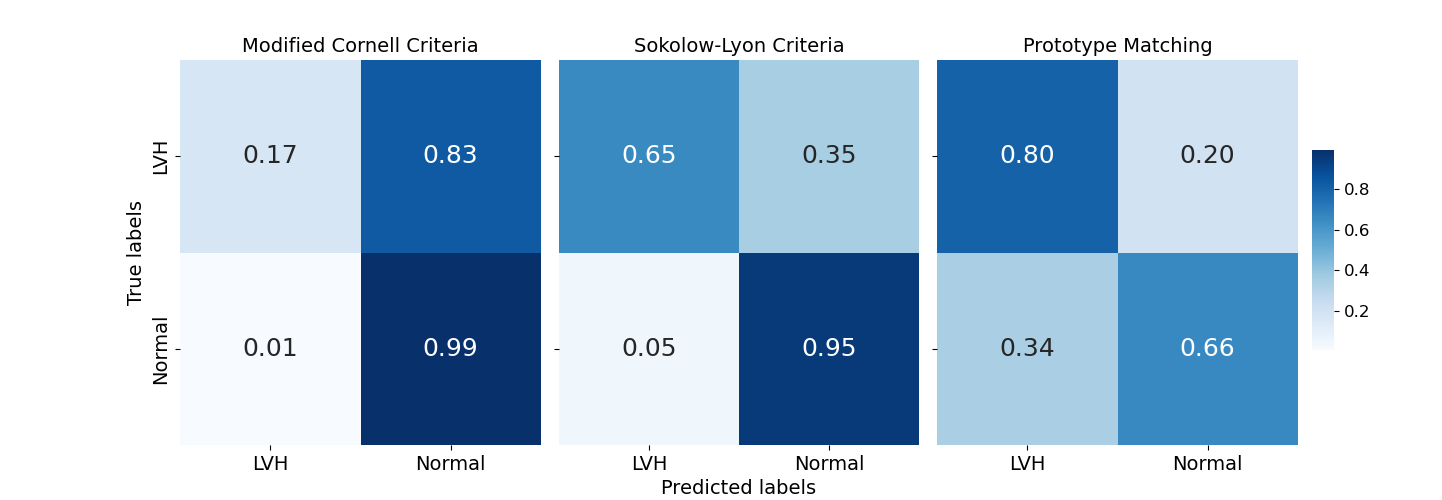}
    \vspace{-0.5cm}
    \caption{Comparison of Normalized Confusion Matrices: We compare our prototype matching method versus two common voltage-based diagnostic criteria for LVH detection.}
    \label{fig:confmatrix}
\end{figure}

\begin{figure}[ht]
    \centering
    
    \begin{subfigure}[t]{\linewidth}
    \centering
    \includegraphics[width = \textwidth]{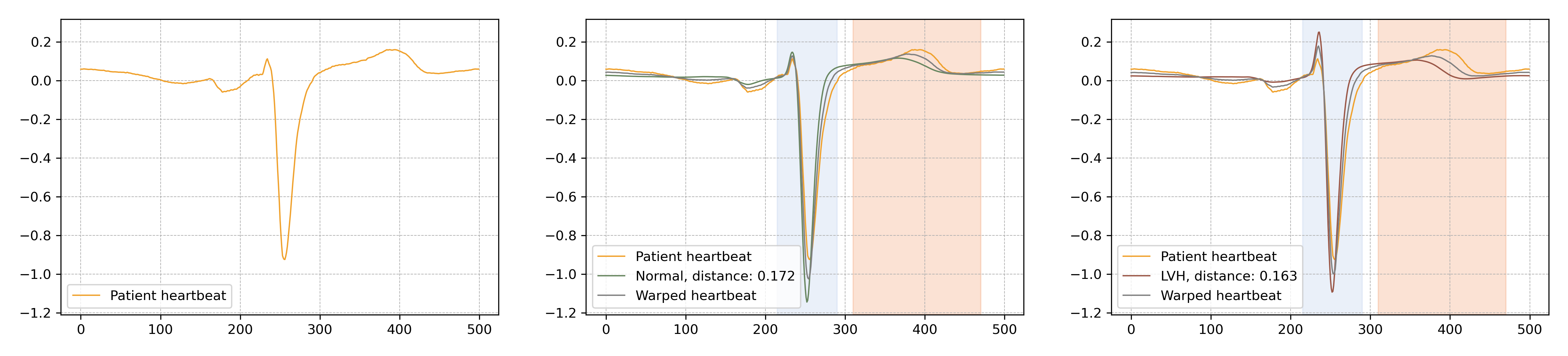} \hspace{0.02in}
    \vspace{-0.5cm}
    \caption{Lead V1}
    \end{subfigure}
    
    \begin{subfigure}[t]{\linewidth}
    \centering
    \includegraphics[width = \textwidth]{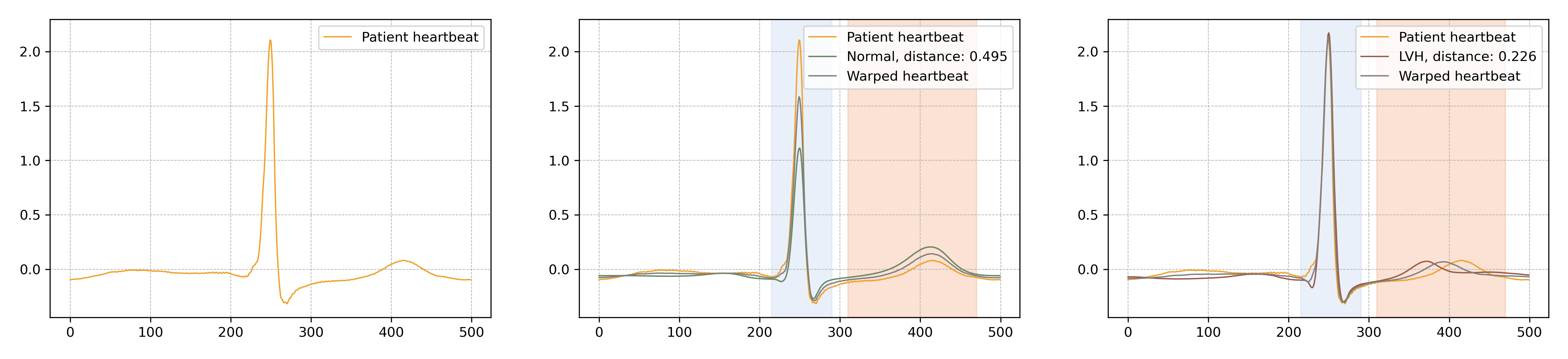} \hspace{0.02in}
    \vspace{-0.5cm}
    \caption{Lead V5}
    \end{subfigure}
    
    \vspace{-0.2cm}
    \caption{A misdiagnosed LVH patient by Sokolow-Lyon criteria: The patient shows a closer proximity of leads V1 and V5 to the LVH prototypes generated by BSW, compared to the normal prototypes.}
    \label{fig:patient}
\end{figure}

\subsection{Data Selection and Preprocessing}

We conducted experiments using the PTB-XL 12-lead ECG dataset \cite{dataset}, which includes 9,069 patients with normal heart function and 456 diagnosed with LVH. Each ECG record is 10 seconds in length, sampled at 500 Hz. We calculated heartbeat variability $v_h$ for each patient, selecting those with $v_h < 0.3$ across all 12 leads. Due to class imbalance, we randomly sampled 256 records from the healthy group to balance against the smaller number of LVH cases, resulting in 215 LVH and 256 normal records for prototype generation.

ECG signals are sensitive to noise, particularly baseline wander, which can impact time warping robustness. To address this, we applied a 4th-order Butterworth high-pass filter to remove baseline drift. For R peak detection, we used NeuroKit2~\cite{Makowski2021neurokit}, customized for our 12-lead data. Given that R peak directions vary across leads, we combined signals from leads \RNum{1}, \RNum{2}, V4, V5, and V6 (which typically have upward R peaks) and subtracted lead aVR (which shows downward R peaks) to enhance R peak prominence. Each heartbeat was segmented from the midpoint between two R peaks to the next, resampled to 500 samples, and averaged within each lead to compute a representative heartbeat, resulting in 12 heartbeat patterns for one patient.

\subsection{Implementation Details}

For BSW (refer to (\ref{eq:loss})), parameters $w_r$ and $w_s$, which regulate the smoothness of $r(t)$ and $s(t)$, are set to $20$ and $10^{-4}$, respectively, with a stricter constraint on $r(t)$ to preserve morphological attributes of each heartbeat. Bounds $s_\text{min}$ and $s_\text{max}$ are set to $-100$ and $100$. The parameter $w_o$, set to $10^{10}$, is triggered only if $s(t)$ falls outside this range, which is uncommon. For affinity calculation, weights for computing the weighted standard deviation of $r(t)$ are derived from the smoothed gradient of the original signals. The threshold for $r(t)$ is dynamically set to $0.015$ times the initial ECG amplitude range, while a fixed threshold of $20$ is applied to $s(t)$.

\subsection{Prototype Validation}

Fig. \ref{fig:warp} presents the application of time warping to two ECG heartbeats. Unlike DTW, the proposed BSW is sensitive to both amplitude and temporal shifts, enabling a more delicate non-linear merging of the signals compared to simple averaging. The resulting prototype libraries are depicted in Fig. ~\ref{fig:comparison}. Notably, prototypes derived from LVH signals often exhibit abnormally high R waves in left-sided leads such as V6 and deep S waves in right-sided leads like V1, indicating that the ventricular depolarization is amplified~\cite{lvh_diagnosis, surawicz2008chou}. In addition, our library also contains some rare prototypes, including those with secondary ST-T changes like inverted T-waves in left-sided leads \RNum{1} and V6, which indicate the presence of a left ventricular “strain” pattern~\cite{brady2020electrocardiogram, T_invert}. These diverse prototypes are preserved to serve as reliable references for accommodating incoming patients with similar atypical symptoms.

To validate the potential applications of the generated prototypes, we employ them as references for diagnosing incoming ECG records. We formulate the distance metric $d$ based on the warping results $r(t)$ and $s(t)$ of an unknown patient's heartbeat pattern relative to the prototypes as:
\[
    d = 10 \times (\left \lVert r(t) - 1\right \rVert_\infty +\sigma(r(t))) + \frac{\left \lVert s(t) \right \rVert_\infty+\sigma(s(t))}{500},
\]
where $\sigma(\cdot)$ denotes the standard deviation. We assign a higher weight to the changes in $r(t)$ as the diagnosis of LVH largely depends on amplitude variations in the waveform. We calculate $d$ to identify the closest two prototypes from both the healthy and LVH libraries per lead, yielding distance sets for each class. Noting that LVH is primarily evident in chest leads V1, V5, and V6, the final decision metric is the summed distances from these leads. If the total distance from the healthy set exceeds that from the LVH set, the patient is classified as having LVH.

The classification was conducted on a test dataset of 100 normal and 100 LVH patients, with confusion matrices compared across various diagnostic criteria (see Fig. \ref{fig:confmatrix}). Compared to the first two criteria, which prioritize reducing false positives but yielding many false negatives, our method reduces these effects by allowing doctors to examine the closest prototype directly. In practice, recognizing that no single automatic method is sufficient by itself, our approach serves as a complementary tool, providing visual support that assists doctors in making reliable final decisions. Fig.~\ref{fig:patient} presents a case misclassified by the Sokolow-Lyon Criteria, where our prototype matching method identified leads V1 and V5 as closer to LVH prototypes.

\section{Conclusion}
\label{sec:conclusion}

This study presents a bilateral signal warping method for generating ECG prototypes that effectively differentiate between LVH and normal heart function. Our approach yields distinct prototypes that align closely with established LVH criteria and can serve as reference standards for diagnosing new ECG records. Experimental results demonstrate that these prototypes enhance LVH diagnosis through prototype matching and also offer valuable support as a diagnostic aid for clinicians. Future research will expand this method to encompass additional cardiac conditions and investigate its applicability to other periodic signals, such as photoplethysmography.

\printbibliography 

@book{brady2020electrocardiogram,
  title={Electrocardiogram in clinical medicine},
  author={Brady, WJ},
  year={2020},
  publisher={John Wiley \& Sons}
}

@article{bacharova2014role,
  title={The role of ECG in the diagnosis of left ventricular hypertrophy},
  author={Bacharova, LME and Schocken, D and H Estes, E and Strauss, D},
  journal={Curr. Cardiol. Rev.},
  volume={10},
  number={3},
  pages={257--261},
  year={2014},
  publisher={Bentham Science Publishers}
}

@article{jothiramalingam2021machine,
  title={Machine learning-based left ventricular hypertrophy detection using multi-lead ECG signal},
  author={Jothiramalingam, R and Jude, A and Patan, R and Ramachandran, M and Duraisamy, JH and Gandomi, AH},
  journal={Neural Comput. Appl.},
  volume={33},
  pages={4445--4455},
  year={2021},
  publisher={Springer}
}

@article{liu2023left,
  title={Left ventricular hypertrophy detection using electrocardiographic signal},
  author={Liu, C and Wu, F and Hu, Y and Pan, R and Lin, C and Chen, Y and Tseng, G and Chan, Y and Wang, C},
  journal={Sci. Rep.},
  volume={13},
  number={1},
  pages={2556},
  year={2023},
  publisher={Nature Publishing Group UK London}
}

@inproceedings{tziakouri2017classification,
  title={Classification of AF and other arrhythmias from a short segment of ECG using dynamic time warping},
  author={Tziakouri, M and Pitris, C and Orphanidou, C},
  booktitle={Proc. CinC},
  pages={1--4},
  year={2017},
  organization={IEEE}
}

@article{schmidt2014two,
  title={Two-dimensional warping for one-dimensional signals—conceptual framework and application to ECG processing},
  author={Schmidt, M and Baumert, M and Porta, A and Malberg, H and Zaunseder, S},
  journal={IEEE Trans. Signal Process.},
  volume={62},
  number={21},
  pages={5577--5588},
  year={2014},
  publisher={IEEE}
}

@article{morel2023time,
  title={Time warping between main epidemic time series in epidemiological surveillance},
  author={Morel, JD and Morel, JM and Alvarez, L},
  journal={PLoS Comput. Biol.},
  volume={19},
  number={12},
  pages={e1011757},
  year={2023},
  publisher={Public Library of Science San Francisco, CA USA}
}

@book{muller2007dynamic,
  author={M{\"u}ller, M},
  title={Information retrieval for music and motion},
  pages={69--84},
  year={2007},
  publisher={Springer}
}

@article{kannel1970electrocardiographic,
  title={Electrocardiographic left ventricular hypertrophy and risk of coronary heart disease: the Framingham Study},
  author={Kannel, WB and Gordon, T and Castelli, WP and Margolis, JR},
  journal={Ann. Intern. Med.},
  volume={72},
  number={6},
  pages={813--822},
  year={1970},
  publisher={American College of Physicians}
}

@article{Makowski2021neurokit,
    author = {D Makowski and T Pham and ZJ Lau and JC Brammer and F Lespinasse and H Pham and C Schölzel and SHA Chen},
    title = {{NeuroKit}2: A Python toolbox for neurophysiological signal processing},
    journal = {Behav. Res. Methods},
    volume = {53},
    number = {4},
    pages = {1689--1696},
    publisher = {Springer Science and Business Media {LLC}},
    doi = {10.3758/s13428-020-01516-y},
    url = {https://doi.org/10.3758%2Fs13428-020-01516-y},
    year = 2021
}

@inproceedings{matchingsurvey,
title={Efficient algorithms for finding maximal matching in graphs},
author={Galil, Z},
booktitle={Proc. CAAP},
pages={90--113},
year={1983},
organization={Springer}
}

@inbook{blossom,
author = {Edmonds, J},
year = {2010},
month = {10},
pages = {361-379},
title = {Path, Trees, and Flowers},
volume = {17},
isbn = {978-0-8176-4841-1},
journal = {Can. J. Math.},
publisher = {Cambridge University Press},
doi = {10.1007/978-0-8176-4842-8_26}
}

@article{dataset,
author = {Wagner, P and Strodthoff, N and Bousseljot, RD and Kreiseler, D and Lunze, F and Samek, W and Schaeffter, T},
year = {2020},
month = {05},
pages = {154},
title = {PTB-XL, a large publicly available electrocardiography dataset},
volume = {7},
journal = {Sci. data},
doi = {10.1038/s41597-020-0495-6}
}

@article{lvh_diagnosis,
author = {Antikainen, R and Grodzicki, T and Palmer, A and Beevers, D and Coles, E and Webster, J and Bulpitt, C},
year = {2003},
month = {03},
pages = {159-64},
title = {The determinants of left ventricular hypertrophy defined by Sokolow-Lyon criteria in untreated hypertensive patients},
volume = {17},
journal = {J. Hum. Hypertens.},
doi = {10.1038/sj.jhh.1001523}
}

@article{T_invert,
author = {Okin, P and Devereux, R and Nieminen, M and Jern, S and Oikarinen, L and Viitasalo, M and Toivonen, L and Kjeldsen, S and Julius, S and Dahlöf, B},
year = {2001},
month = {08},
pages = {514-520},
title = {Relationship of the electrocardiographic strain pattern to left ventricular structure and function in hypertensive patients: The LIFE study},
volume = {38},
journal = {J. Am. Coll. Cardiol.},
doi = {10.1016/S0735-1097(01)01378-X}
}

@article{AI1,
author = {Rudin, C},
year = {2019},
month = {05},
pages = {206-215},
title = {Stop Explaining Black Box Machine Learning Models for High Stakes Decisions and Use Interpretable Models Instead},
volume = {1},
journal = {Nat. Mach. Intell.},
doi = {10.1038/s42256-019-0048-x}
}

@inproceedings{AI2,
author = {Baylor, E and Beede, E and Hersch, F and Iurchenko, A and Ruamviboonsuk, P and Vardoulakis, L and Wilcox, L},
year = {2020},
month = {01},
title = {A Human-Centered Evaluation of a Deep Learning System Deployed in Clinics for the Detection of Diabetic Retinopathy},
booktitle = {Proc. CHI},
doi = {10.1145/3313831.3376718}
}

@article{AI_lvh,
author = {Wu, J and Tsai, M and Xiao, S and Liaw, Y},
year = {2020},
month = {03},
title = {A deep neural network electrocardiogram analysis framework for left ventricular hypertrophy prediction},
journal = {J. Ambient Intell. Humaniz. Comput.},
doi = {10.1007/s12652-020-01826-1}
}

@article{AI_lvh2,
author = {Yu, X and Yao, X and Wu, B and Zhou, H and Xia, S and Su, W and Wu, Y and Zheng, X},
year = {2021},
month = {11},
title = {Using deep learning method to identify left ventricular hypertrophy on echocardiography},
volume = {38},
journal = {Int. J. Cardiovasc. Imaging},
doi = {10.1007/s10554-021-02461-3}
}

@book{surawicz2008chou,
  title={Chou's electrocardiography in clinical practice: adult and pediatric},
  author={Surawicz, B and Knilans, T},
  year={2008},
  publisher={Elsevier Health Sciences}
}

@inproceedings{interp_ai,
author = {Li, O and Liu, H and Chen, C and Rudin, C},
title = {Deep learning for case-based reasoning through prototypes: a neural network that explains its predictions},
year = {2018},
isbn = {978-1-57735-800-8},
booktitle = {Proc. AAAI}
}

@article{interp_ai2,
author = {Barnett, A and Schwartz, F and Tao, C and Chen, C and Ren, Y and Lo, J and Rudin, C},
year = {2021},
month = {12},
title = {A case-based interpretable deep learning model for classification of mass lesions in digital mammography},
volume = {3},
journal = {Nat. Mach. Intell.},
doi = {10.1038/s42256-021-00423-x}
}

@inproceedings{tang2024optimized,
  title={Optimized Hard Exudate Detection with Supervised Contrastive Learning},
  author={Tang, W and Cui, K and Chan, RH},
  booktitle={Proc. ISBI},
  pages={1--5},
  year={2024},
  organization={IEEE}
}

@article{eeg,
author = {Barnett, A and Guo, Z and Jing, J and Ge, W and Kaplan, P and Kong, W and Karakis, I and Herlopian, A and Jayagopal, L and Taraschenko, O and Selioutski, O and Osman, G and Goldenholz, D and Rudin, C and Westover, MB},
year = {2024},
month = {05},
title = {Improving Clinician Performance in Classifying EEG Patterns on the Ictal–Interictal Injury Continuum Using Interpretable Machine Learning},
volume = {1},
journal = {NEJM AI},
doi = {10.1056/AIoa2300331}
}

\end{document}